\documentclass[preprint,aps]{revtex4-1}
\usepackage{hyperref}
\usepackage{graphicx}
\usepackage{epstopdf}
\usepackage{amsmath}
\def\be{\begin{equation}}
\def\ee{\end{equation}}
\def\bea{\begin{eqnarray}}
\def\eea{\end{eqnarray}}
\def\ba{\begin{array}}
\def\ea{\end{array}}
\def\bc{\begin{center}}
\def\ec{\end{center}}
\begin{document}
\title{Single Transverse Spin Asymmetries in Semi-inclusive Deep Inelastic Scattering in a Spin-1 Diquark Model }
\author{Narinder Kumar and Harleen Dahiya}
\affiliation{Department of Physics\\
Dr. B. R. Ambedkar National Institute of Technology\\
         Jalandhar-144011, India}
\begin{abstract}
The observed results for the azimuthal single spin asymmetries (SSAs) of the proton, measured in the semi-inclusive deep inelastic scattering (SIDIS), can be explained by the final-state interaction (FSI) from the gluon exchange between the outgoing quark and the target spectator system. SSAs require a phase difference between two amplitudes coupling the target with opposite spins to the same final state. We have used the model of light front wave functions (LFWFs) consisting of a spin-$\frac{1}{2}$ system as a composite of a spin-$\frac{1}{2}$ fermion and a spin-1 vector boson to estimate the SSAs.
The implications of such a model have been investigated in detail by considering different coupling constants.  The FSIs also produce a complex phase which can be included in the LFWFs to calculate the Sivers and Boer-Mulders distribution functions of the nucleon.
\end{abstract}
\maketitle
\section{Introduction}
In recent years, there has been lot of interest to investigate the single spin asymmetries (SSAs) \cite{dennis,dennis1} which are considered the most interesting phenomenon to understand Quantum Chromodynamics (QCD) from the basic principles. When the target nucleon is transversely polarized to an incoming beam, many particles (hadrons) are produced in collision which tend to show right-left asymmetry in their distribution relative to nucleon spin direction. One of the participating particles in scattering process carries a polarization and if the scattering cross section depends on the direction of this polarization, a SSA is obtained. One of the remarkable aspect about the SSAs is that they persist at high energy and show a very stable pattern. A strong correlation between the target proton spin $\vec{S}_p$ and the plane of the produced pion and virtual photon has been observed in the semi-inclusive deep inelastic scattering (SIDIS) experiments by the HERMES and SMC collaborations \cite{hermes,hermes1,smc}. COMPASS, a fixed target experiment at CERN \cite{compass,compass1,compass2,compass3,compass4}, presents the result for the SSA of charged hadrons produced in deep inelastic scattering of muons on a transversely polarized target. STAR collaboration \cite{star,star1,star2} shows that the SSA for $p p \rightarrow \pi X$ process at center of mass energies of 20 and 200 GeV  may arise from the Sivers effect \cite{dennis}, Collins effect \cite{collins} or a combination of both.
A large azimuthal SSA was observed in the hadronic reactions like $p p^\uparrow \rightarrow \pi X$ \cite{E704} and in $p p \rightarrow \Lambda^\uparrow X$ \cite{spin96} where the antiproton and hyperon respectively are polarized normal to the production plane. The issue regarding SSAs becomes all the more difficult to understand when it is realized that the sign of the SIDIS experimental and Drell-Yan (DY) experimental \cite{drellyan} measurements are different.

The fact that the understanding of SSAs theoretically depends simultaneously on several aspects of hadron structure, for example, interference between complex phases, orbital angular momentum and final state interactions (FSI))\cite{brodsky,drellyan,ellis,corelator1}, makes the study of SSAs even more interesting.
It was shown in Ref. \cite{burkardt} that a relation exists between the generalized parton distributions (GPDs) and SSAs (SSAs= FSI $\times$ GPDs). In particular, the transverse distortion of the distribution of quarks in a transversely polarized target nucleon is related to the spin flip GPD $E(x,0,t)$. It was shown that FSIs from gluon exchange between the outgoing quark and the target lead to SSAs \cite{brodsky} which essentially require a complex phase difference between the two amplitudes which couple the proton target with $J_p^z=\pm\frac{1}{2}$ to the final state. The SSAs have been studied for both the SIDIS and DY process by using a QCD motivated quark-diquark model of a spin-$\frac{1}{2}$ proton of mass $M$ with spin-$\frac{1}{2}$ and spin-0 constituents of mass $m$ and $\lambda$ respectively \cite{brodsky2}. SSA for DY process was explained with the help of initial state interactions (ISI) \cite{drellyan,boer}.  Collins \cite{collins} pointed out important consequences of SSAs in deep inelastic scattering (DIS) by incorporating the struck quark into Wilson line path order exponents. The theoretical description of the single spin phenomena in QCD is still a big challenge.

A comprehensive picture of the nucleon can been obtained by considering the transverse momentum dependent parton distributions (TMDs). TMDs are the T-odd parton distributions which give rise to SSAs and describe various hard exclusive reactions of the nucleon \cite{diehl_report,radyushkin}. The FSIs can also act as a source of the Sivers and Boer-Mulders distribution functions \cite{Yuan,belitsky,boer1}. Sivers distribution function $(f_{1 T}^{\perp})$, proposed by Sivers in 1990 \cite{dennis}, describes the distribution of unpolarized quarks inside the nucleons which are transversely polarized in the opposite direction. Due to PT invariance Sivers asymmetry was supposed to be prohibited $f_{1 T}=-f_{1 T}=0$ \cite{collins}. However, Brodsky, Hwang and Schmidt proved that $f_{1 T} \ne 0$ \cite{brodsky}. On the other hand, Boer-Mulders function $(h_{1}^{\perp})$ describes the distribution of polarized quarks inside the unpolarized nucleons. Extensive amount of work has been done in the light cone quark models \cite{boffi,pasquini,pasquini1} as well as diquark models which include the inclusive pion and kaon production in DIS \cite{ellis}. Sivers distribution function has also been discussed in the MIT bag model with the help of interference of $S$ and $P$ wave components in the presence of FSIs \cite{fyuan}. Further, model independent and dependent relations between the GPDs and TMDs have also been developed \cite{metz}. The derivatives of chiral odd GPDs in the impact parameter space can be related  with $f_{1 T}^{\perp}$  and $h_{1}^{\perp}$  as well as with the spin densities of the nucleon \cite{bpasquini}.
The SSA and the Sivers distribution function can be related on the basis of the interference terms coming from the FSI. Recently, Sivers and Boer-Mulders distribution function have been studied on the lattice to non-local operators \cite{musch}.  We would like to emphasize that the signs of $f_{1 T}^{\perp}$  and $h_{1}^{\perp}$ are of great interest as different models predict different signs for them. Therefore, it becomes interesting to study the Sivers and Boer-Mulders distribution function in the context of SSAs and FSIs.

The set of light front wave functions (LFWFs) provide a frame-independent, quantum-mechanical description of hadrons at the amplitude level which are capable of encoding multi-quark and gluon momentum, helicity and flavor correlations in the form of universal process independent hadron wave functions. One can also construct the invariant mass operator $H_{LC}= P^+ P^- - P_\perp^2$ and light cone time operator $P^-=P^0-P^z$  in the light cone gauge from the QCD Lagrangian \cite{rev,wf,wf1}. Since the essential physics that allows the SSA is the existence of two different light front angular momentum components in the proton $L_z=0$, $L_z=\pm 1$ with different FSIs, therefore it becomes interesting to calculate the SSA in the SIDIS  process $\gamma^* p \rightarrow q (qq)_1$ where a virtual photon is scattered on a transversely polarized proton having transverse spin and is induced by the FSIs. In this model, the spin-$\frac{1}{2}$ proton of mass $M$ is a composite of spin-$\frac{1}{2}$ fermion of mass $m$ and spin-1 vector boson of mass $\lambda$ and it lies in the framework of QED \cite{model,hadron_optics,impact,kumar_dahiya1,kumar_dahiya2,chiralgpds}. This model has been successfully used as a template for calculating the Schwinger anomalous magnetic moment, understanding the structure of relativistic composite systems and their matrix elements in hadronic physics, explaining the hadronic structure and providing a good representation of 3-D structure of the nucleon \cite{hadron_optics,impact,chiralgpds}. Further, it also gives the general proof for the vanishing of the anomalous gravitomagnetic moment $B(0)$ \cite{wf,kumar_dahiya1,kumar_dahiya2}.  This model has been used to calculate the spin and orbital angular momentum of a composite relativistic system as well as the GPDs in impact parameter space \cite{chakrabarti1,chakrabarti2}.

In the present work, we have calculated the single spin asymmetry (SSA) in the semi-inclusive deep inelastic scattering (SIDIS) process $\gamma^* p \rightarrow q (qq)_1$. The amplitude for this process is computed at both the tree and one-loop level and the SSA has been calculated by taking interference between these amplitudes. Final-state interaction (FSI) from the gluon exchange between the outgoing quark and the target spectator system has been used to explain the observed results for the azimuthal SSAs of the proton. SSAs require a phase difference between two amplitudes coupling the target with opposite spins to the same final state. To estimate the SSAs, we have used the model of light front wave functions (LFWFs) consisting of a spin-$\frac{1}{2}$ proton of mass $M$ as a composite of a spin-$\frac{1}{2}$ fermion of mass $m$ and a spin-1 vector boson of mass $\lambda$. The FSIs also produce a complex phase which can be included in the LFWFs to calculate the Sivers and Boer-Mulders distribution functions of the nucleon. Sivers function is given by the overlap of the wave functions having opposite proton spin states but the same quark spin state whereas the Boer-Mulders function has the same proton spin state but opposite quark spin
state. The Sivers and Boer-Mulders distribution functions have been computed by inducing the spin-dependent complex phases to the LFWFs. The implications of such a model have been investigated in detail by considering fixed and running coupling constants.

\section{light front wave functions  (LFWFs)}
\begin{figure}
\minipage{0.42\textwidth}
    \includegraphics[width=7cm]{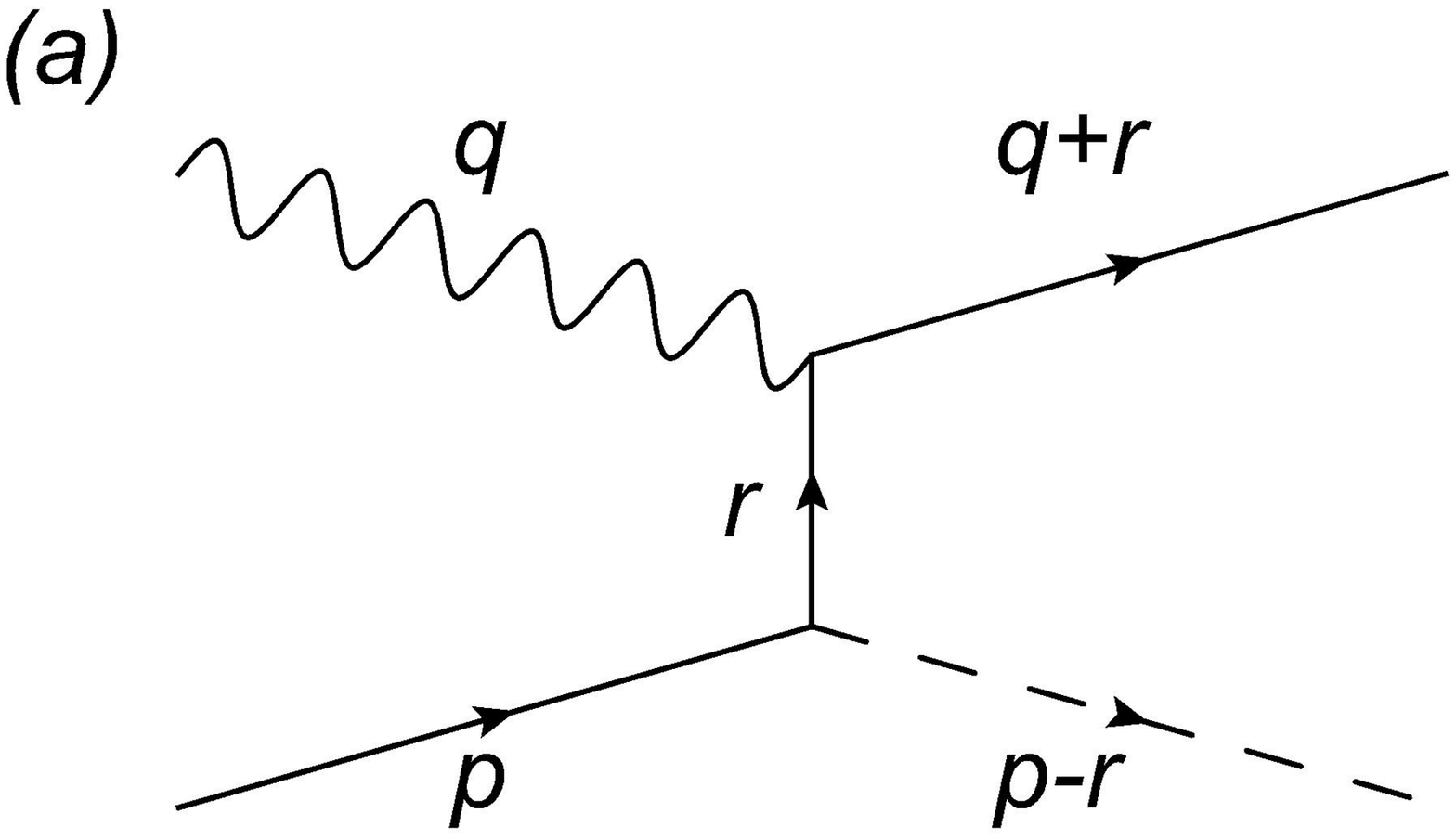}
  \endminipage\hfill
  \minipage{0.42\textwidth}
  \includegraphics[width=7cm]{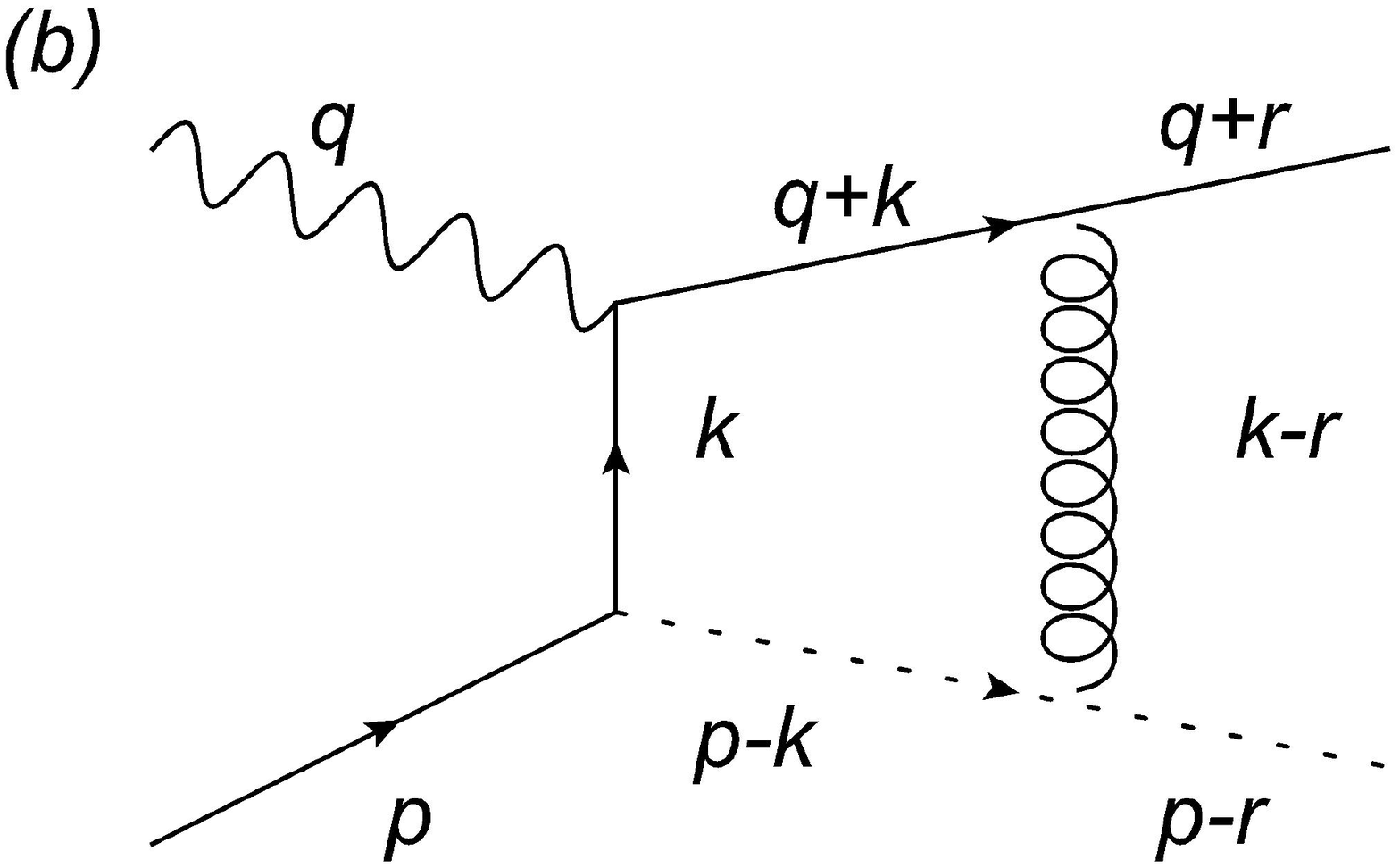}
\endminipage\hfill
 \caption{The (a) tree diagram and (b) one-loop for $\gamma^* p \rightarrow q (qq)_1$. The interference amplitudes obtained from these two diagrams provides the SSA.}
  \label{fsdiagram}
\end{figure}

The single spin asymmetry is calculated in the SIDIS process $\gamma^* p \rightarrow q (qq)_1$ where virtual photon is scattered with virtuality $q^2=- Q^2$ on a transversely polarized proton having transverse spin where a spin-$\frac{1}{2}$ system acts as a composite of a spin-$\frac{1}{2}$ fermion and a spin-1 vector boson.
During the process, the virtual photon is absorbed by an active spin-$\frac{1}{2}$ quark  as shown in Fig. \ref{fsdiagram} and the diquark is used to describe the spectator system.  The tree level diagram (Fig. \ref{fsdiagram} (a)) alone will not describe the observed asymmetry because the contribution is real towards the scattering amplitude. However, the interference between the tree level and the one-loop (Fig. \ref{fsdiagram} (b)) amplitudes together will generate the  desired SSA.
Kinematics have been developed in the Drell-Yan-West frame which is boosted so that $q^+=0$ and is collinear to proton \cite{brodsky}. The longitudinal momentum fraction exchanged is defined in terms of the momentum carried by the outgoing quark $r$ and the momentum carried by the photon $q$ as $\Delta= \frac{r^+}{p^+}$. We can fix $r^-$ and $q^-$ from the momentum conservation and the on-shell conditions for the nucleon, quark and diquark and we have
\bea
r^-&=& \frac{M^2}{p^+}- \frac{\vec{r}_\perp^2 + \lambda^2}{(1- \Delta)p^+},\nonumber\\
q^- &\approx& \frac{Q^2+2 \vec{q}_\perp \cdot \vec{r}_\perp}{\Delta p^+}.
\eea
The other kinematical details can be summarized as follows
\bea
p^\mu &=& \Big(p^+, \frac{M^2}{p^+},~ 0_\perp \Big),\nonumber\\
q^\mu &=& \Big(0, \frac{(\vec{q}_\perp + \vec{r}_\perp)^2}{\Delta p^+}- \frac{M^2}{p^+}-\frac{\vec{r}_\perp^2+\lambda^2}{(1-\Delta)p^+},~\vec{q}_\perp\Big),\nonumber\\
r^\mu &=& \Big( \Delta p^+, \frac{M^2}{p^+}-\frac{\vec{r}_\perp^2+\lambda^2}{(1-\Delta)p^+}, ~\vec{r}_\perp \Big).
\eea
We consider here a spin-$\frac{1}{2}$ system which acts a composite of spin-$\frac{1}{2}$ fermion and spin-1 vector boson \cite{brodsky,wf1,chakrabarti1,chakrabarti2}. If we consider $x$ as the fraction of momentum transferred and ${\vec k}_{\perp}$ as the transverse component of the momentum, the two-particle Fock state for a fermion with $J^z = + \frac{1}{2}$, having four possible spin combinations, can be expressed as
\begin{eqnarray}
&&\left|\Psi^{\uparrow}_{\rm two \ particle}(P^+, \vec P_\perp = \vec
0_\perp)\right>
\ =\
\int\frac{{\mathrm d}^2 {\vec k}_{\perp} {\mathrm d} x }{{\sqrt{x(1-x)}}16
\pi^3}
\label{vsn1}\\
&\times&
\Big[ \
\psi^{\uparrow}_{+\frac{1}{2}\, +1}(x,{\vec k}_{\perp})\,
\left| +\frac{1}{2}\, +1\, ;\,\, xP^+\, ,\,\, {\vec k}_{\perp}\right>
+\psi^{\uparrow}_{+\frac{1}{2}\, -1}(x,{\vec k}_{\perp})\,
\left| +\frac{1}{2}\, -1\, ;\,\, xP^+\, ,\,\, {\vec k}_{\perp}\right>
\nonumber\\
&&\ \ \ +\psi^{\uparrow}_{-\frac{1}{2}\, +1} (x,{\vec k}_{\perp})\,
\left| -\frac{1}{2}\, +1\, ;\,\, xP^+\, ,\,\, {\vec k}_{\perp}\right>
+\psi^{\uparrow}_{-\frac{1}{2}\, -1} (x,{\vec k}_{\perp})\,
\left| -\frac{1}{2}\, -1\, ;\,\, xP^+\, ,\,\, {\vec k}_{\perp}\right>\ \Big]
\ ,
\nonumber
\end{eqnarray}
where
\begin{eqnarray}
&&\psi_{+\frac{1}{2}+1}^{\uparrow}(x,\vec{k}_\perp)=-\sqrt{2}\frac{-k^1+i k^2}{x(1-x)}\varphi,\nonumber\\ && \psi_{+\frac{1}{2}-1}^{\uparrow}(x,\vec{k}_\perp)=-\sqrt{2}\frac{k^1+ ik^2}{(1-x)}\varphi,\nonumber\\ &&
\psi_{-\frac{1}{2}+1}^{\uparrow}(x,\vec{k}_\perp)=-\sqrt{2}\left(M-\frac{m}{x}\right)\varphi,\nonumber\\ &&\psi_{-\frac{1}{2}-1}^{\uparrow}(x,\vec{k}_\perp)=0. \
\label{spinup}
\end{eqnarray}
Similarly, for $J^z=- \frac{1}{2}$ we  have four possible combinations
\begin{eqnarray}
&&\left|\Psi^{\downarrow}_{\rm two \ particle}(P^+, \vec P_\perp =
\vec 0_\perp)\right>
\ =\
\int\frac{{\mathrm d}^2 {\vec k}_{\perp} {\mathrm d} x }{{\sqrt{x(1-x)}}16
\pi^3}
\label{vsn1a}\\
&\times&
\Big[\
\psi^{\downarrow}_{+\frac{1}{2}\, +1}(x,{\vec k}_{\perp})\,
\left| +\frac{1}{2}\, +1\, ;\,\, xP^+\, ,\,\, {\vec k}_{\perp}\right>
+\psi^{\downarrow}_{+\frac{1}{2}\, -1}(x,{\vec k}_{\perp})\,
\left| +\frac{1}{2}\, -1\, ;\,\, xP^+\, ,\,\, {\vec k}_{\perp}\right>
\nonumber\\
&&\ \ \ +\psi^{\downarrow}_{-\frac{1}{2}\, +1}(x,{\vec k}_{\perp})\,
\left| -\frac{1}{2}\, +1\, ;\,\, xP^+\, ,\,\, {\vec k}_{\perp}\right>
+\psi^{\downarrow}_{-\frac{1}{2}\, -1}(x,{\vec k}_{\perp})\,
\left| -\frac{1}{2}\, -1\, ;\,\, xP^+\, ,\,\, {\vec k}_{\perp}\right>\ \Big]
\ ,
\nonumber
\end{eqnarray}
where
\begin{eqnarray}
&&\psi_{+\frac{1}{2}+1}^{\downarrow}(x,\vec{k}_\perp)=0,\nonumber\\&&
\psi_{+\frac{1}{2}-1}^{\downarrow}(x,\vec{k}_\perp)=-\sqrt{2}\left(M-\frac{m}{x}\right)\varphi,\nonumber\\ &&\psi_{-\frac{1}{2}+1}^{\downarrow}(x,\vec{k}_\perp)=-\sqrt{2}\frac{-k^1+i k^2}{(1-x)}\varphi,\nonumber\\
&&\psi_{-\frac{1}{2}-1}^{\downarrow}(x,\vec{k}_\perp)=-\sqrt{2}\frac{k^1+i k^2}{x(1-x)}\varphi \,,
\label{spindown}
\end{eqnarray}
and
\begin{eqnarray}
\varphi(x, \vec{k}_{\perp}) =\frac{e}{\sqrt {1-x}} \frac{1}{M^2-\frac{\vec{k}^2_{\perp}+m^2}{x}-\frac{\vec{k}_{\perp}^{2}+\lambda^2}{1-x}}\,.
\end{eqnarray}
Here the formalism has been generalized by assigning a mass $M$ to external fermions in the scattering process but a different mass $m$ to the internal fermion line and a mass $\lambda$ to the internal boson line. The charge of the fermion is taken as $e$. It may be important to mention here that the numerators in the wave functions are characteristic of the orbital angular momentum and hold for both perturbative and non-perturbative couplings.

\section{Tree and one loop amplitudes}

The SSA can be calculated from the interference of the amplitudes at tree  and one-loop levels as shown in Fig. \ref{fsdiagram}.
The tree level amplitude, calculated from Fig. \ref{fsdiagram} (a), receives contribution from the following amplitudes with an $\mathcal{O}(\alpha_s)$ FSI
\bea
\mathcal{A}^{Tree}(\Uparrow \rightarrow \uparrow, s_b^z=+1)&=& - \sqrt{2} \frac{(-r^1+i r^2) C}{\Delta (1-\Delta)(r_\perp^2-M^2 \Delta (1- \Delta)+m^2 (1-\Delta)+\lambda^2 \Delta)},\nonumber\\
\mathcal{A}^{Tree}(\Uparrow \rightarrow \uparrow, s_b^z=-1)&=& - \sqrt{2} \frac{(r^1+i r^2) C}{(1-\Delta) (r_\perp^2- M^2 \Delta (1- \Delta)+m^2 (1-\Delta)+\lambda^2 \Delta)},\nonumber\\
\mathcal{A}^{Tree}(\Downarrow \rightarrow \uparrow, s_b^z=+1)&=& - \sqrt{2} \Big(M-\frac{m}{\Delta}\Big) \frac{C}{(r_\perp^2- M^2 \Delta (1- \Delta)+m^2 (1-\Delta)+\lambda^2 \Delta)},\nonumber\\
\mathcal{A}^{Tree}(\Downarrow \rightarrow \uparrow, s_b^z=-1)&=& - \sqrt{2} \Big(M-\frac{m}{\Delta}\Big) \frac{C}{(r_\perp^2- M^2 \Delta (1- \Delta)+m^2 (1-\Delta)+\lambda^2 \Delta)},\nonumber\\
\mathcal{A}^{Tree}(\Downarrow \rightarrow \downarrow, s_b^z=+1)&=& - \sqrt{2} \frac{(-r^1+i r^2) C}{\Delta (1-\Delta)(r_\perp^2-M^2 \Delta (1- \Delta)+m^2 (1-\Delta)+\lambda^2 \Delta)},\nonumber\\
\mathcal{A}^{Tree}(\Downarrow \rightarrow \downarrow, s_b^z=-1)&=& - \sqrt{2} \frac{(r^1+i r^2) C}{(1-\Delta) (r_\perp^2- M^2 \Delta (1- \Delta)+m^2 (1-\Delta)+\lambda^2 \Delta)},\label{tree-final}
\eea
where
\be
C= - e \ e_1 P^+ \sqrt{\Delta} \ 2 \ \Delta \ (1-\Delta). \label{c}
\ee
We have taken here the electric charges of $q$ and $(qq)_1$ to be $e_1$ and $e_2$. The state $\Uparrow (\Downarrow)$ denote the two spin states of the proton $J_p^z= \pm \frac{1}{2}$ whereas for the spin projection of the spin-$\frac{1}{2}$ constituent $J_q^z=\pm \frac{1}{2}$ we label $\uparrow$ and $\downarrow$.

The one-loop amplitudes can be expressed as
\bea
\mathcal{A}^{one-loop}(I)&=&i e e_1^{2} e_2 \int \frac{d^4 k \ \mathcal{N}(I)}{(k^2-m^2+i \epsilon)((k+q)^2-m^2+i \epsilon)} \times \nonumber\\
&& \frac{1}{((k-r)^2-\lambda_g^2+i \epsilon)((k-p)^2-\lambda^2+i \epsilon)},
\label{oneloop_amplitude}
\eea
where $\lambda_g$ is the mass of the gauge boson. In the standard leading order perturbative QCD (pQCD) calculations \cite{pqcd},  final state multiple interactions are considered whereas in the present calculation the spin asymmetry is obtained by considering a single final state interaction. The spin asymmetry in pQCD is obtained from the non-vanishing transverse momentum dependent cross section and in our calculations we consider the process $\gamma^* p \rightarrow q (qq)_1$ where the detected particle is identical to the quark and the asymmetry for this detected hadron can be obtained by convoluting the jet asymmetry result with the realistic fragmentation function. A similarity in both the approaches is that a required phase is essential to generate the asymmetry and this phase comes from a pole in the propagator.

The numerators $\mathcal{N}(I)$ in Eq. (\ref{oneloop_amplitude}) can be obtained from the Feynman diagram (Fig. \ref{fsdiagram} (b)) as follows
\bea
\mathcal{N}(\Uparrow \rightarrow \uparrow, (s_b^z=+1))&=& 2 P^+ \sqrt{\Delta} \ x \Big(-\sqrt{2}\  \frac{-k^1+ i k^2}{x(1-x)}\Big) q^-, \nonumber\\
\mathcal{N}(\Uparrow \rightarrow \uparrow, (s_b^z=-1))&=& 2 P^+ \sqrt{\Delta} \ x \Big(-\sqrt{2}\ \frac{k^1+ i k^2}{1-x}\Big) q^- ,\nonumber\\
\mathcal{N}(\Uparrow \rightarrow \downarrow, (s_b^z=+1))&=& 2 P^+ \sqrt{\Delta} \  x \Big(-\sqrt{2}\ \Big(M-\frac{m}{x})\Big)q^-,\nonumber\\
\mathcal{N}(\Downarrow \rightarrow \uparrow, (s_b^z=-1))&=& 2 P^+ \sqrt{\Delta} \ x \Big(-\sqrt{2} \ \Big(M-\frac{m}{x})\Big)q^-,\nonumber\\
\mathcal{N}(\Downarrow \rightarrow \uparrow, (s_b^z=+1))&=& 2 P^+ \sqrt{\Delta} \ x \Big(-\sqrt{2} \ \frac{-k^1+ i k^2}{1-x}\Big)q^- ,\nonumber\\
\mathcal{N}(\Downarrow \rightarrow \downarrow, (s_b^z=-1))&=& 2 P^+ \sqrt{\Delta} \ x \Big(-\sqrt{2}\  \frac{k^1+ i k^2}{x(1-x)}\Big)q^-.
\label{numerator_term}
\eea
where $q^-= \frac{Q^2}{\Delta P^+}$ with $\Delta=\frac{k^+}{P^+}$ being the quark light cone fraction. The denominators in Eq. (\ref{oneloop_amplitude}) can be rewritten as
\bea
k^2-m^2+i \epsilon &=& k^+\Big(k^- - \frac{k_\perp^2+m^2-i \epsilon}{k^+}\Big),\nonumber\\
(k+q)^2-m^2+i \epsilon &=&  (k^+ + q^+) \Big((k^-+ q^-)- \frac{m^2+(k_\perp+q_\perp)^2-i \epsilon}{x P^+}\Big),\nonumber\\
(k-r)^2-\lambda_g^2+i \epsilon&=& (k^+- r^+)\Big((k^- - r^-)- \frac{\lambda_g^2+ (k_\perp-r_\perp)^2-i \epsilon}{(x-\Delta) P^+}\Big),\nonumber\\
(k-p)^2-\lambda^2+i \epsilon&=& (k^+- P^+) \Big((k^- - P^-) + \frac{\lambda^2 + k_\perp^2 - i \epsilon}{(1-x) P^+}\Big),
\eea
whereas the integral over the momentum space in light front field theory can be expressed as
\be
\int d^4 k=\int \frac{d^2 k_\perp dk^+ dk^-}{2(2 \pi)^4}.
\ee
The one-loop amplitude now becomes
\bea
\mathcal{A}^{one-loop}(I)&=& - i e e_1^2 e_2 \int \frac{d^2 k_\perp}{2(2 \pi)^4} \int \frac{P^+ \mathcal{N}(I) dx}{(P^+)^4 x^2 (x-\Delta) (1-x)} \times \nonumber\\
&& \int \frac{dk^-}{\Big(k^- - \frac{k_\perp^2+m^2-i \epsilon}{x P^+}\Big)\Big((k^-+ q^-)- \frac{m^2+(k_\perp+q_\perp)^2-i \epsilon}{x P^+}\Big)}\times \nonumber\\
&& \frac{1}{\Big((k^- - r^-)- \frac{\lambda_g^2+ (k_\perp-r_\perp)^2-i \epsilon}{(x-\Delta) P^+}\Big)\Big((k^- - P^-) + \frac{\lambda^2 + k_\perp^2 - i \epsilon}{(1-x) P^+}\Big)}. \label{oneloop}
\eea
We would like to mention here that there is a restriction on the  combination of propagators that can go on-shell simultaneously and the cuts have to be imposed by hand. These kinematical constraints have been discussed in detail in Ref. \cite{brodsky2}. On performing the integration over $k^-$, we get
\bea
\mathcal{A}^{one-loop}(I) &=& -i e e_1^2 e_2 (2 \pi i) \int \frac{d^2 k_\perp}{2(2 \pi)^4}\int P^+ \frac{\mathcal{N}(I) dx}{(P^+)^4 x^2 (x-\Delta) (1-x)} \times \nonumber\\
&& \frac{1}{\Big(P^- - \frac{\lambda^2 + k_\perp^2 - i \epsilon}{(1-x) P^+} - \frac{m^2 + k_\perp^2 - i \epsilon}{x P^+}\Big) \Big(P^- + q^- -  \frac{\lambda^2 + k_\perp^2 - i \epsilon}{(1-x) P^+} - \frac{m^2+(k_\perp+q_\perp)^2-i \epsilon}{x P^+} \Big)} \times \nonumber\\
&& \frac{1}{\Big(P^- - r^- - \frac{\lambda^2 + k_\perp^2 - i \epsilon}{(1-x) P^+}- \frac{\lambda_g^2+(k_\perp-r_\perp)^2-  i \epsilon}{(x- \Delta)P^+} \Big)}.
\label{amp}
\eea
The imaginary part of the above equation gives rise to the imaginary phase which is essential for the SSAs. This imaginary part evolves from the real intermediate propagator state allowed just before re-scattering and can be defined in terms of the initial state ($P_{init}^{-}$) and intermediate state ($P_{interm}^{-}$) energy denominators as
\be
Im \Big(\frac{1}{P_{init}^{-}- P_{interm}^{-}+ i \epsilon}\Big)=- \pi \ \delta(P_{init}^{-}- P_{interm}^{-}).
\ee
The imaginary part from the propagator term from Eq. (\ref{amp}) gives
\bea
\frac{1}{P^-+q^- - \frac{\lambda^2+\vec{k}_\perp^2 -i\epsilon}{(1-x)P^+}-\frac{m^2+(\vec{k}_\perp+\vec{q}_\perp)^2-i \epsilon}{x P^+}} &=& -i\pi \delta\Big(P^-+q^- - \frac{\lambda^2+\vec{k}_\perp^2}{(1-x)P^+}-\frac{m^2+(\vec{k}_\perp+\vec{q}_\perp)^2}{x P^+}\Big)\nonumber\\&=& - i \pi \frac{\Delta^2 P^+}{\vec{q}_\perp^2} \delta(x-\Delta- \bar{\delta}),
\label{imaginary_part}
\eea
where
\be
\bar{\delta}= 2 \Delta \frac{\vec{q}_\perp \cdot (\vec{k}_\perp - \vec{r}_\perp)}{\vec{q}^2_\perp}.
\ee
In terms of $\bar{\delta}$ the  exchanged momentum $\bar{\delta}P^+$ is very small. This leads to the light cone energy denominator being dominated by $\frac{(\vec{k}_\perp-\vec{r}_\perp)^2+\lambda_g^2}{x-\Delta}$ term.

Using Eqs. (\ref{numerator_term}), (\ref{amp}),and (\ref{imaginary_part}), one-loop amplitudes obtained can be written as
\bea
\mathcal{A}^{one-loop}(\Uparrow \rightarrow \uparrow, (s_b^z=+1))&=& - i e e_1^2 e_2 \int \frac{d^2 k_\perp 2 P^+ \sqrt{\Delta} \ \Delta \ (1-\Delta) \sqrt{2} (k^1-i k^2)}{16 \pi^2 \Delta(1-\Delta) L_1 L_2},\nonumber\\
\mathcal{A}^{one-loop}(\Uparrow \rightarrow \uparrow, (s_b^z=-1))&=& i e e_1^2 e_2 \int \frac{d^2 k_\perp 2 P^+ \sqrt{\Delta} \ \Delta \ (1-\Delta) \sqrt{2} (k^1+i k^2)}{16 \pi^2 (1-\Delta)  L_1 L_2},\nonumber\\
\mathcal{A}^{one-loop}(\Uparrow \rightarrow \downarrow, (s_b^z=+1))&=& i e e_1^2 e_2 \int \frac{d^2 k_\perp 2 P^+ \sqrt{\Delta} \ \Delta \ (1-\Delta) \sqrt{2} (M-\frac{m}{\Delta})}{ L_1 L_2}, \nonumber\\
\mathcal{A}^{one-loop}(\Downarrow \rightarrow \uparrow, (s_b^z=-1))&=& i e e_1^2 e_2 \int \frac{d^2 k_\perp 2 P^+ \sqrt{\Delta} \ \Delta \ (1-\Delta) \sqrt{2} (M-\frac{m}{\Delta})}{ L_1 L_2}, \nonumber\\
\mathcal{A}^{one-loop}(\Downarrow \rightarrow \downarrow, (s_b^z=-1))&=& - i e e_1^2 e_2 \int \frac{d^2 k_\perp 2 P^+ \sqrt{\Delta} \ \Delta \ (1-\Delta) \sqrt{2} (k^1-i k^2)}{16 \pi^2 (1-\Delta) L_1 L_2},\nonumber\\
\mathcal{A}^{one-loop}(\Downarrow \rightarrow \uparrow, (s_b^z=+1))&=& i e e_1^2 e_2 \int \frac{d^2 k_\perp 2 P^+ \sqrt{\Delta} \ \Delta \ (1-\Delta) \sqrt{2} (k^1+i k^2)}{16 \pi^2 \Delta(1-\Delta)  L_1 L_2}, \label{oneloop-final}
\eea
where
\bea
L_1&=& \vec{k}_\perp^2- M^2 \Delta (1-\Delta)+m^2(1-\Delta)+ \lambda^2 \Delta, \nonumber\\
L_2&=& \lambda_g^2+(\vec{k}_\perp- \vec{r}_\perp)^2.
\eea
These can now be integrated over the transverse momentum $k_\perp$ using Feynman parametrization. Using Eqs. (\ref{tree-final}) and (\ref{oneloop-final}), the interference between the tree level and one-loop amplitudes can be calculated and we get
\bea
\mathcal{A}(\Uparrow \rightarrow \uparrow, (s_b^z=+1))&=& \frac{- \sqrt{2} (-r^1+i r^2)}{\Delta(1-\Delta)} C \Big(h+ i \frac{e_1 e_2}{16 \pi^2}I_2\Big),\nonumber\\
\mathcal{A}(\Uparrow \rightarrow \uparrow, (s_b^z=-1))&=& \frac{- \sqrt{2} (r^1+i r^2)}{1-\Delta} C \Big(h+ i \frac{e_1 e_2}{16 \pi^2}I_2\Big),\nonumber\\
\mathcal{A}(\Uparrow \rightarrow \downarrow, (s_b^z=+1))&=& - \sqrt{2} \Big(M- \frac{m}{\Delta}\Big) C \Big(h+ i \frac{e_1 e_2}{16 \pi^2}I_1\Big),\nonumber\\
\mathcal{A}(\Downarrow \rightarrow \uparrow, (s_b^z=-1))&=& - \sqrt{2} \Big(M- \frac{m}{\Delta}\Big) C \Big(h+ i \frac{e_1 e_2}{16 \pi^2}I_1\Big),\nonumber\\
\mathcal{A}(\Downarrow \rightarrow \downarrow, (s_b^z=+1))&=& \frac{- \sqrt{2} (-r^1+i r^2)}{1-\Delta} C \Big(h+ i \frac{e_1 e_2}{16 \pi^2}I_2\Big),\nonumber\\
\mathcal{A}(\Downarrow \rightarrow \uparrow, (s_b^z=+1))&=& \frac{- \sqrt{2} (r^1+i r^2)}{\Delta(1-\Delta)} C \Big(h+ i \frac{e_1 e_2}{16 \pi^2}I_2\Big),
\label{interference}
\eea
where $C$ has already been defined in Eq. (\ref{c}) and
\bea
I_1&=& \int_{0}^{1} d\alpha \frac{1}{D},\nonumber\\
I_2&=&\int_{0}^{1} d\alpha \frac{\alpha}{D},\nonumber\\
h&=& \frac{1}{\vec{r}_\perp^2 - M^2 \Delta (1-\Delta)+m^2 (1-\Delta)+\lambda^2 \Delta},\nonumber\\
D&=&\alpha (1-\alpha)\  \vec{r}_\perp^2+\alpha \ \lambda_g^2- M^2 (1-\alpha) \Delta (1-\Delta)+m^2 (1-\alpha)(1-\Delta)+ \lambda^2 (1-\alpha) \Delta.
\eea
\section{Single spin asymmetry (SSA)}
We now present the results for the SSA. Following Refs. \cite{brodsky,brodsky2} we have used the gauge particle as an infrared regulator and have taken the mass of gauge boson as $\lambda_g=0$. The calculations have been performed in the region $\Delta < x < 1$. The production plane has been taken as the $\hat{z}-\hat{x}$ plane which can be defined from the virtual photon and the produced hadron. The asymmetry is produced in the $\hat{y}$ direction and can be expressed as
\bea
\mathcal{P}_y &=& \frac{1}{\mathcal{C}} \Big(i (\mathcal{A}(\Uparrow \rightarrow \uparrow)^* \mathcal{A}(\Downarrow \rightarrow \uparrow);(s_b^z=+1)- \mathcal{A}(\Uparrow \rightarrow \uparrow) \mathcal{A}(\Downarrow \rightarrow \uparrow)^*;(s_b^z=+1)) + \nonumber\\
&& i (\mathcal{A}(\Uparrow \rightarrow \downarrow)^* \mathcal{A}(\Downarrow \rightarrow \downarrow);(s_b^z=+1)-\mathcal{A}(\Uparrow \rightarrow \downarrow) \mathcal{A}(\Downarrow \rightarrow \downarrow)^*;(s_b^z=+1))+ \nonumber\\
&& i (\mathcal{A}(\Uparrow \rightarrow \uparrow)^*, \mathcal{A}(\Downarrow \rightarrow \uparrow);(s_b^z=-1)-\mathcal{A}(\Uparrow \rightarrow \uparrow) \mathcal{A}(\Downarrow \rightarrow \uparrow)^*;(s_b^z=-1))+\nonumber\\
&& i (\mathcal{A}(\Uparrow \rightarrow \downarrow)^* \mathcal{A}(\Downarrow \rightarrow \downarrow);(s_b^z=-1)-\mathcal{A}(\Uparrow \rightarrow \downarrow) \mathcal{A}(\Downarrow \rightarrow \downarrow)^*;(s_b^z=-1))\Big),
\eea
where
\bea
\mathcal{C}&=&|\mathcal{A}(\Uparrow \rightarrow \uparrow,(s_b^z= +1))|^2+|\mathcal{A}(\Uparrow \rightarrow \uparrow,(s_b^z= -1))|^2+|\mathcal{A}(\Uparrow \rightarrow \downarrow,(s_b^z= +1))|^2+\nonumber\\
&& |\mathcal{A}(\Downarrow \rightarrow \uparrow,(s_b^z= -1))|^2+|\mathcal{A}(\Downarrow \rightarrow \downarrow,(s_b^z= +1))|^2+|\mathcal{A}(\Downarrow \rightarrow \downarrow,(s_b^z= -1))|^2.
\eea
Substituting the amplitudes from Eq. (\ref{interference}), we get azimuthal spin asymmetry transverse to production plane as follows
\bea
\mathcal{P}_y&=& - \frac{e_1 e_2}{8 \pi} \frac{r^1 (M \Delta-m)\Delta (1-\Delta) (\vec{r}_\perp^2- M^2 \Delta (1-\Delta)+m^2 (1-\Delta)+ \lambda^2 \Delta)}{(\vec{r}_\perp^2(1+\Delta^2)+(M \Delta - m)^2 (1-\Delta)^2)}\times \nonumber\\
&& \frac{1}{\vec{r}_\perp^2}ln\frac{\vec{r}_\perp^2- M^2 \Delta (1-\Delta)+m^2 (1-\Delta)+ \lambda^2 \Delta)}{- M^2 \Delta (1-\Delta)+m^2 (1-\Delta)+ \lambda^2 \Delta}.
\label{final_asymmetry}
\eea
Here the factor $r^1=r^x$ reflects that single spin asymmetry is proportional to the term $\vec{S}_p \cdot \vec{q} \times \vec{r}$.

\begin{figure}
\minipage{0.42\textwidth}
    \includegraphics[width=7cm]{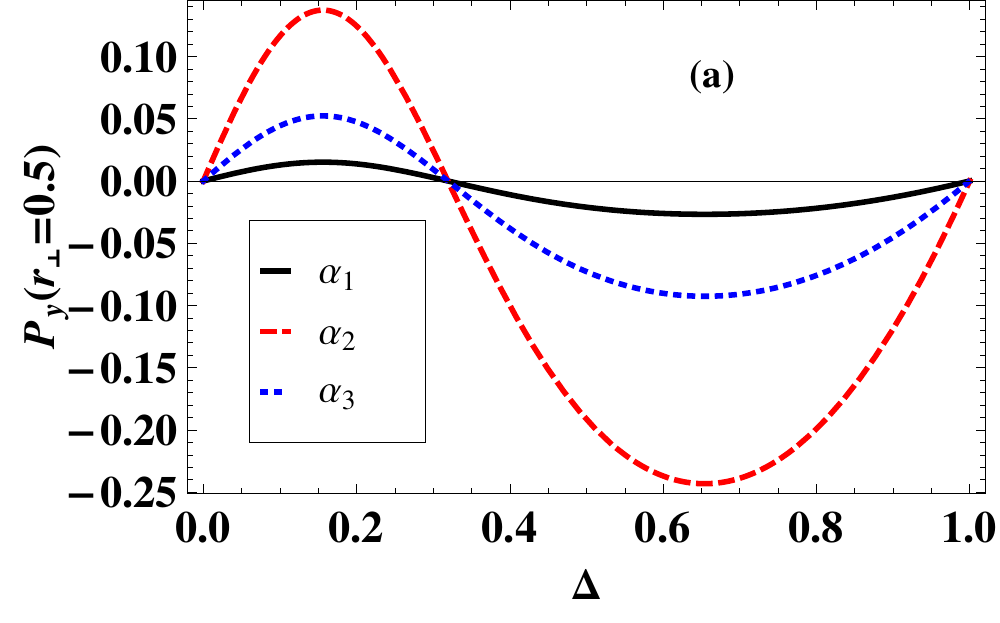}
  \endminipage\hfill
  \minipage{0.42\textwidth}
  \includegraphics[width=7cm]{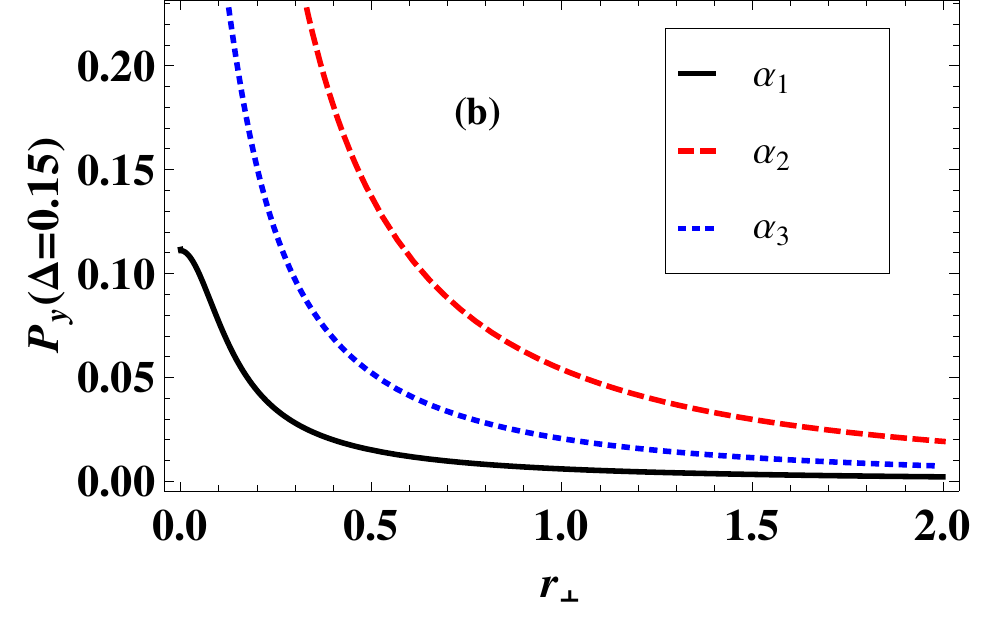}
\endminipage\hfill
 \caption{Transverse asymmetry $\mathcal{P}_y$ as a function of $\Delta$ and $r_\perp$ of the proton obtained by gluon exchange in FSI.}
  \label{asymmetry}
\end{figure}

The FSI from gluon exchange in the present case has the strength $\frac{e_1 e_2}{4 \pi} \rightarrow C_F \alpha_1(Q^2)$, where $C_F$ is taken to be $\frac{4}{3}$ so that the results match with QCD. Momentum transfer carried by the gluon fixes the scale for the strong coupling constant $\alpha_1$ which has a fixed value of $\alpha_1=0.3$ as obtained in $\overline{MS}$ scheme \cite{brodsky1}. In the present work the masses of the nucleon, quark, diquark and gluon have respectively be taken as $M=0.94$ GeV, $m=0.3$ GeV, $\lambda=0.8$ GeV and $\lambda_g=0$ GeV \cite{brodsky}. For the sake of comparison as well as a deeper understanding of the role of coupling constant we have also calculated the asymmetry by replacing the fixed coupling constant $(\alpha_1)$ by  running coupling constants where higher order contributions, particularly the closed quark loops,  can be taken into account. We take an analytic one-loop running coupling constant as proposed by the Shirkov and Sovlovstov \cite{baldichhi1,baldichhi2,shirkov}
\be
\alpha_2(Q^2)=\frac{4 \pi}{\beta_0} \frac{1}{\ ln(Q^2/\Lambda_{QCD}^2)},
\label{shirkov}
\ee
with $\beta_0=11-\frac{2}{3} N_f$ ($N_f$ being the number of active quarks). The use of such running coupling constant was found to be essential for explaining the reproduction of the light pseudoscalar meson spectrum. However, for $\Lambda_{QCD}^2=Q^2$, an nonphysical singularity existed which contradicted some analytical properties and had to be modified in the infrared region. Therefore, the above said coupling constant was modified \cite{baldichhi1,baldichhi2,shirkov} such that it remains finite at $\Lambda_{QCD}^2=Q^2$ and is given by
\be
\alpha_3(Q^2)= \frac{4 \pi}{\beta_0}\Big(\frac{1}{ln(Q^2/\Lambda_{QCD}^2)}+\frac{\Lambda_{QCD}^2}{\Lambda_{QCD}^2-Q^2}\Big).
\label{shirkov_a}
\ee
In order to check the behaviour and dependence of asymmetries on light cone momentum fraction $\Delta$ and momentum carried by the outgoing quark $r_\perp$, in Fig. \ref{asymmetry} we have presented the model prediction for the transverse azimuthal spin asymmetry  $\mathcal{P}_y$ as a function $\Delta$ and $r_\perp$ for the fixed coupling constant $\alpha_1=0.3$ and the running coupling constants, $\alpha_2$ and $\alpha_3$. 

\begin{figure}
\minipage{0.42\textwidth}
    \includegraphics[width=7cm]{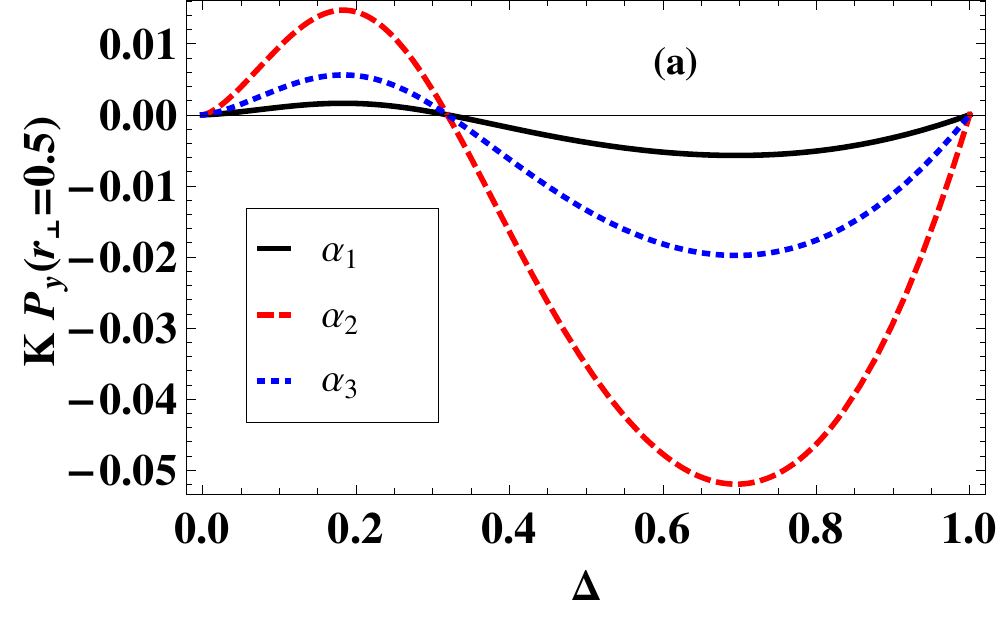}
  \endminipage\hfill
  \minipage{0.42\textwidth}
  \includegraphics[width=7cm]{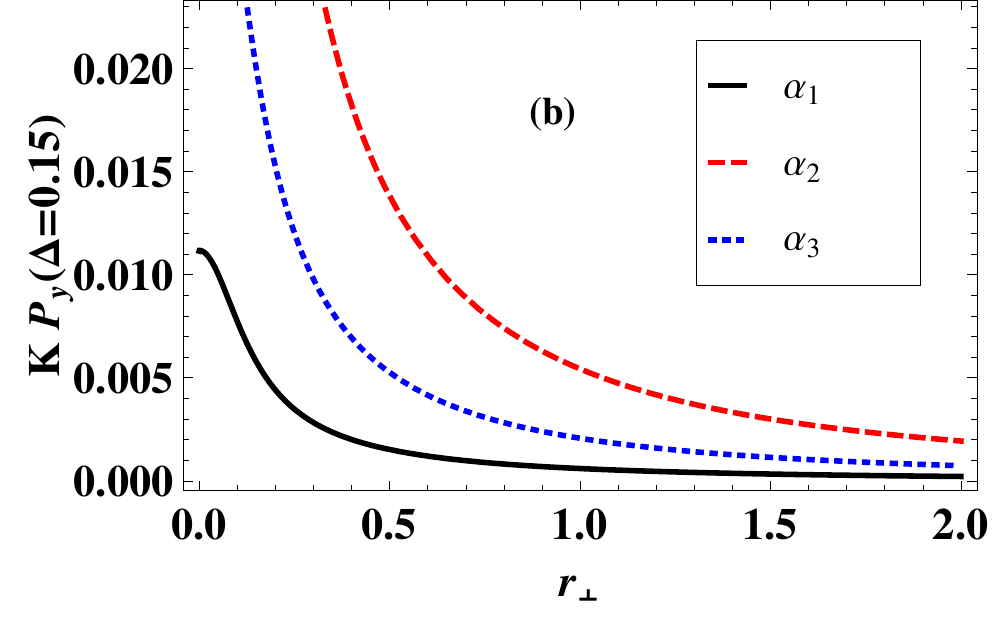}
\endminipage\hfill
\caption{Longitudinal asymmetry $K \mathcal{P}_y$ as a function of $\Delta$ and $r_\perp$ of the proton obtained by gluon exchange in FSI.}
  \label{asymmetry_hermes}
\end{figure}

The longitudinal azimuthal spin asymmetry can be determined from the transverse azimuthal spin asymmetry $\mathcal{P}_y$ using the relation
\be
A_{U L}^{\sin\phi}=K A_{U T}^{\sin\phi},
\ee
where $A_{U T}^{\sin\phi}$ is HERMES transverse asymmetry where the target polarization is transverse to the incident lepton direction, $A_{U L}^{\sin\phi}$ is the HERMES longitudinal asymmetry where the target nucleon is polarized along the incident lepton direction \cite{hermes}. The kinematic factor $K$ is defined in terms of the virtual photon fractional energy $y$ as
\be
K= \sqrt{\frac{2 M x}{E_{lab}}} \sqrt{\frac{1-y}{y}}.
\ee
We have taken $y=0.5$ and $E_{lab}=27.6$  GeV \cite{brodsky_polonica}. The longitudinal azimuthal spin asymmetry $K \mathcal{P}_y$ is plotted in Fig. \ref{asymmetry_hermes} as a function $\Delta$ and $r_\perp$ for all the three coupling constants mentioned above.

In Figs. \ref{asymmetry} (a) and \ref{asymmetry_hermes} (a) we have shown the variation of transverse and longitudinal asymmetries ($\mathcal{P}_y$ and $K \mathcal{P}_y$) with $\Delta$ at a fixed value of momentum carried by the outgoing quark $r_\perp=0.5$ GeV. Interestingly, the transverse as well as the longitudinal asymmetries change sign at $\Delta \sim 0.3$ as the value of the $\Delta$ increases. This may be possibly due to the $q\bar{q}$ pairs being readily produced at this momentum fraction which give rise to sea quarks affecting the sign of asymmetry at this point. The sea quarks seem to play an important role in carrying the momentum fraction. Future experiments like E906/SeaQuest Drell-Yan experiment \cite{paul_reimer,nakahara} are ready to study the sign change of asymmetries at the momentum fraction $\Delta \sim 0.3$. Further, in Figs. \ref{asymmetry} (b) and \ref{asymmetry_hermes} (b) we have fixed the momentum fraction $\Delta=0.15$ GeV and plotted $\mathcal{P}_y$ and $K \mathcal{P}_y$ as a function of quark transverse momentum $r_\perp$. Since $r_\perp$ is the transverse momentum of the outgoing quark relative to photon direction, the asymmetries do not change sign in these cases. It is also clear from the Eq. (\ref{final_asymmetry}) that $\mathcal{P}_y$ decreases as $\alpha_1 \ \Delta \ \frac{1}{r_\perp^2} ln \ r_\perp^2$. The values of $Q^2$ and $\Lambda^2$ in the above calculations are taken to be 0.25 GeV and 0.10 GeV respectively.

\section{Sivers and Boer-Mulders distribution functions}
Quark transverse momentum dependent parton distributions  are defined through the correlation function as follows \cite{corelator,sivers_efermov,sivers_burkardt}
\bea
\Phi^{q[\Gamma]}(x,\vec{k}_\perp,S)=\frac{1}{2}\int \frac{dy^-}{2 \pi} \frac{d^2 y_\perp}{(2 \pi)^2} e^{i k \cdot y} \langle P,S| \bar{\psi}(0)\Gamma \mathcal{W}\psi(y)|P,S\rangle |_{y^+=0}.
\eea
Here $P$ is the nucleon momentum, $x$ and $\vec{k}_\perp$ are the fractional longitudinal and the transverse momentum of the quark, $S$ is the covariant spin vector having components $S^+=\frac{S_L P^+}{M}$, $S^-=-\frac{S_L P^-}{M}$ and $ \vec{S}_\perp$. The gauge link (Wilson line) $\mathcal{W}$ assures color gauge invariance of the correlators and $\Gamma$ are the usual gamma matrices. By projecting the correlators onto the full basis of $\gamma$ matrices one can obtain the corresponding distribution functions. At leading twist the proton spin structure  can be described by 8 TMDs. Using the gamma matrices $\gamma^+$, $\gamma^+ \gamma_5$ and $i \sigma^{j+}\gamma_5$, we obtain
\bea
\Phi^{[\gamma^+]}(x,\vec{k}_\perp,S)&=& f_1(x,{\vec k}_{\perp})- \frac{\epsilon^{i j}_{\perp}k^{i}_\perp S^{j}_\perp}{M} f^{\perp}_{1T}(x,{\vec k}_{\perp}),\nonumber\\
\Phi^{[\gamma^+ \gamma_5]}(x,\vec{k}_\perp,S)&=& S_L g_{1L}(x,{\vec k}_{\perp})+ \frac{\vec{k}_\perp \cdot \vec{S}_\perp}{M} g_{1T}(x,{\vec k}_{\perp}),\nonumber\\
\Phi^{[i \sigma^{+ j}\gamma_5]}(x,\vec{k}_\perp,S)&=& \frac{\epsilon^{i j}_{\perp}k^{i}_\perp }{M}h^{\perp}_{1}(x,{\vec k}_{\perp})+ \frac{\lambda k^{j}_{\perp}}{M} h^{\perp}_{1L}(x,{\vec k}_{\perp})+ S^{j}_\perp \Big( h_{1T}(x,{\vec k}_{\perp})+ \frac{\vec{k}_\perp^2}{2 M^2}h^{\perp}_{1T}(x,{\vec k}_{\perp})\Big)+ \nonumber\\
&&\frac{2 k^{j}_{\perp} \vec{k}_\perp \cdot \vec{S}_\perp- S^{j}_{\perp} \vec{k}_\perp^2}{2 M^2} h^{\perp}_{1 T}.
\eea
The LFWF representation of the Sivers ($f^{\perp}_{1T}(x,{\vec k}_{\perp})$) and Boer-Mulders ($h^{\perp}_{1}(x,{\vec k}_{\perp})$) distribution functions can be deduced from the proton eigensolution $|\psi_p \rangle$ on the eigenstates $\{ |n\rangle\}$ of the free Hamiltonian giving the light cone Fock expansion as
\bea
|\psi_p (P^+, \vec{P_\perp})\rangle &=& \sum_{n, \lambda_i}
\prod_{i=1}^{n}
{{\rm d}x_{i}\, {\rm d}^2{\vec{k}}_{\perp i} \over 16\pi^3 }\ \,
16\pi^3 \delta\left(1-\sum_{j=1}^n x_j\right) \, \delta^{(2)}
\left(\sum_{j=1}^n {\vec{k}}_{\perp j}\right) \nonumber\\
&& \times \psi_n (x_i, \vec{k}_{\perp i}, \lambda_i) |n; x_i P^+, x_i \vec{P}_\perp+ \vec{k}_{\perp i}, \lambda_i \rangle.
\eea
The relative momentum coordinates of the LFWFs are the momentum fractions $x_i= \frac{k^{+}_{i}}{P^+}$ and the transverse momenta $\vec{k}_{\perp i}$ of partons.  The light cone spin projections along the direction of quantization are expressed as $\lambda_i$ whereas the physical transverse momenta of partons are $\vec{p}_{\perp i}= x_i \vec{P}_\perp +\vec{k}_{\perp i}$ \cite{sivers_hwang}.

The distribution functions $f_1(x,{\vec k}_{\perp})$ and $f^{\perp}_{1T}(x,{\vec k}_{\perp})$ can be defined through the matrix elements of the bilinear vector current as follows
\begin{eqnarray}
&&\int\frac{d y^-d^2{\vec y}_{\perp}}{16\pi^3}\;
e^{ix P^+y^--i{\vec k}_{\perp}\cdot {\vec y}_{\perp}}\;
\langle P,{\vec S}_{\perp} | \bar\psi(0)\,\gamma^+\,\psi(y)\,|P,{\vec S}_{\perp}
\rangle
\Big|_{y^+=0}
\label{siv1}\\
&&=
{1\over 2 P^+}\ \Big[\ f_1(x,{\vec k}_{\perp})\ {\bar U}(P,{\vec S}_{\perp})
\ {\gamma^+} \ U(P,{\vec S}_{\perp})
\ +\
f_{1T}^{\perp}(x,{\vec k}_{\perp})\ {k_{\perp}^i\over M}\ {\bar U}(P,{\vec S}_{\perp})
\ \sigma^{i+} \ U(P,{\vec S}_{\perp}) \ \Big]
\ ,
\nonumber
\end{eqnarray}
where
\begin{equation}
{1\over 2 P^+}
{\bar U}(P,{\vec S}_{\perp}) \sigma^{i+} U(P,{\vec S}_{\perp})=
\epsilon^{ji}S_{\perp}^j\ \qquad
{\rm with}\ \qquad \epsilon^{12}=-\epsilon^{21}=1\ .
\label{siv2}
\end{equation}
For the present work, we consider the transverse spin $\vec{S}_\perp= (S_\perp^1,S_\perp^2)=(0,1)$. The proton state can be defined as a combination of two states as $\frac{(|P, \uparrow \rangle+ i | P, \downarrow \rangle)}{\sqrt{2}}$ and we have
\bea
&&\int\frac{d y^-d^2{\vec y}_{\perp}}{16\pi^3}\;
e^{ix P^+y^--i{\vec k}_{\perp}\cdot {\vec y}_{\perp}} {\langle P,\uparrow | -i \langle P,\downarrow | \over {\sqrt{2}}}
\bar\psi(0)\,\gamma^+\,\psi(y)\,
{|P,\uparrow \rangle +i |P,\downarrow \rangle \over {\sqrt{2}}}
\Big|_{y^+=0}
 \nonumber\\
&& =f_1(x,{\vec k}_{\perp}) - S_{\perp}^2\ {k_{\perp}^1\over M}\ f_{1T}^{\perp}(x,{\vec k}_{\perp}).
\eea
We can now write
\bea
f_1(x,{\vec k}_{\perp})&=& \int\frac{d y^-d^2{\vec y}_{\perp}}{16\pi^3}\;
e^{ix P^+y^--i{\vec k}_{\perp}\cdot {\vec y}_{\perp}} {1\over 2}\;\Big[\langle P,\uparrow |J^+(y)|P,\uparrow \rangle
+ \langle P,\downarrow |J^+(y)|P,\downarrow \rangle \Big]
\Big|_{y^+=0}, \nonumber\\
-{k_{\perp}^1\over M}\ f_{1T}^{\perp}(x,{\vec k}_{\perp})&=& {i\over 2}\;\Big[\langle P,\uparrow |J^+(y)|P,\downarrow \rangle
- \langle P,\downarrow |J^+(y)|P,\uparrow \rangle \Big]
\Big|_{y^+=0},
\label{sivers_bm}
\eea
with $J^+(y)= \bar{\psi}(0) \gamma^+ \psi(y)$.

The Boer-Mulders distribution function can be defined through the matrix elements of the bilinear tensor current as follows
\begin{eqnarray}
&&\int\frac{d y^-d^2{\vec y}_{\perp}}{16\pi^3}\;
e^{ix P^+y^--i{\vec k}_{\perp}\cdot {\vec y}_{\perp}}\;
\langle P,{\vec S}_{\perp} | \, \bar\psi(0)\,\sigma^{i+}\,\psi(y)\,
|P,{\vec S}_{\perp}
\rangle
\Big|_{y^+=0}
\label{siv1h1p} \nonumber \\
&&\qquad =\
{1\over 2 P^+}\ \Big[\ h_{1}^{\perp}(x,{\vec k}_{\perp})\
{k_{\perp}^i\over M}\ {\bar U}(P,{\vec S}_{\perp})
\ \gamma^{+} \ U(P,{\vec S}_{\perp}) \ \Big],
\end{eqnarray}
leading to
\be
{k_{\perp}^i\over M}\ h_{1}^{\perp}(x,{\vec k}_{\perp})={1\over 2}
\int\frac{d y^-d^2{\vec y}_{\perp}}{16\pi^3}\;
e^{ix P^+y^--i{\vec k}_{\perp}\cdot {\vec y}_{\perp}}\
\Big( \langle P,\uparrow |J^{\sigma^{i+}}(y)|P,\uparrow \rangle+\langle P,\downarrow |J^{\sigma^{i+}}(y)|P,\downarrow \rangle \Big)
\Big|_{y^+=0},
\label{boerm}
\ee
with
$J^{\sigma^{i+}}(y)= \bar{\psi}(0) \sigma^{i+} \psi(y)$.

The light cone representation of unpolarized quark distribution and Sivers function can be written from Eq. (\ref{sivers_bm}) and is expressed as
\begin{equation}
f_1(x,{\vec k}_{\perp})
={\cal C'}\
{1\over 2}\ \Big[ \psi^{\uparrow \ *}_{(n)}(x_i,
  {\vec{k}}_{\perp i},\lambda_i) \
\psi^{\uparrow}_{(n)}(x_i, {\vec{k}}_{\perp i},\lambda_i)
\ +\
\psi^{\downarrow \ *}_{(n)}(x_i,
  {\vec{k}}_{\perp i},\lambda_i) \
\psi^{\downarrow}_{(n)}(x_i, {\vec{k}}_{\perp i},\lambda_i)
\Big] \ ,
\label{p11sa}
\end{equation}
\begin{equation}
-{k_{\perp}^1\over M}\ f_{1T}^{\perp}(x,{\vec k}_{\perp})
=
{\cal C'}\
{i\over 2}\ \Big[ \psi^{\uparrow \ *}_{(n)}(x_i,
  {\vec{k}}_{\perp i},\lambda_i) \
\psi^{\downarrow}_{(n)}(x_i, {\vec{k}}_{\perp i},\lambda_i)
\ -\
\psi^{\downarrow \ *}_{(n)}(x_i,
  {\vec{k}}_{\perp i},\lambda_i) \
\psi^{\uparrow}_{(n)}(x_i, {\vec{k}}_{\perp i},\lambda_i)
\Big] \ ,
\label{sivers}
\end{equation}
where
\begin{equation}
{\cal C'}\ \equiv \
\sum_{n, \lambda_i}
\int \prod_{i=1}^{n}
{{\rm d}x_{i}\, {\rm d}^2{\vec{k}}_{\perp i} \over 16\pi^3 }\ \,
16\pi^3 \delta\left(1-\sum_{j=1}^n x_j\right) \, \delta^{(2)}
\left(\sum_{j=1}^n {\vec{k}}_{\perp j}\right)\
\delta(x-x_{1})\ \delta^{(2)}({\vec{k}}_{\perp}-{\vec{k}}_{\perp 1})\ .
\label{pp1}
\end{equation}
On the other hand, the light cone representation of the Boer-Mulders function can be written from Eq. (\ref{boerm}) and is expressed as

\begin{eqnarray}
{k_{\perp}^1\over M}\ h_{1}^{\perp}(x,{\vec k}_{\perp})
&=&
{{\cal C'}\over 2}\ (-i)\
\Big(
\Big[ \ \psi^{\uparrow \ *}_{(n)}(x_i,
  {\vec{k}}_{\perp i},{\lambda}^{\prime}_1=\downarrow ,\lambda_{i\ne 1})
\ \psi^{\uparrow}_{(n)}(x_i, {\vec{k}}_{\perp i},\lambda_1=\uparrow ,\lambda_{i\ne 1})
\nonumber\\
&&\qquad\ \ \ -\
\psi^{\uparrow \ *}_{(n)}(x_i,
  {\vec{k}}_{\perp i},{\lambda}^{\prime}_1=\uparrow ,\lambda_{i\ne 1})
\ \psi^{\uparrow}_{(n)}(x_i, {\vec{k}}_{\perp i},\lambda_1=\downarrow ,\lambda_{i\ne 1})
\ \Big]
\nonumber\\
&&\ \
\ \ \ \ \ \ +\
\Big[ \ \psi^{\downarrow \ *}_{(n)}(x_i,
  {\vec{k}}_{\perp i},{\lambda}^{\prime}_1=\downarrow ,\lambda_{i\ne 1})
\ \psi^{\downarrow}_{(n)}(x_i, {\vec{k}}_{\perp i},\lambda_1=\uparrow ,\lambda_{i\ne 1})
\nonumber\\
&&\qquad\ \ \ -\
\psi^{\downarrow \ *}_{(n)}(x_i,
  {\vec{k}}_{\perp i},{\lambda}^{\prime}_1=\uparrow ,\lambda_{i\ne 1})
\ \psi^{\downarrow}_{(n)}(x_i, {\vec{k}}_{\perp i},\lambda_1=\downarrow ,\lambda_{i\ne 1})
\ \Big]
\Big) \ . \ \ \ \ \ \
\label{boer_mulders}
\end{eqnarray}
It is clear from Eq. (\ref{sivers}) that the Sivers function is the product of LFWFs having opposite proton spin states but the same quark spin state whereas for the Boer-Mulders function from Eq. (\ref{boer_mulders}) we note that LFWFs have same proton spin state and opposite quark spin state.

For the calculations of Sivers and Boer-Mulders distribution functions in the model of spin-1 diquark model, we adopt the same treatment for the LFWFs as in the earlier section.  We have induced spin-dependent complex phases to the wave functions which can be expressed as
\bea
\psi^\uparrow_{+\frac{1}{2}+1}(x,\vec{k}_\perp)&=& \frac{- \sqrt{2}(-k^1+ i k^2)}{x(1-x)}\left(1+ i \frac{e_1 e_2}{8 \pi} (\vec{k}_\perp^2+B) I_4\right) \varphi ,\nonumber\\
\psi^\uparrow_{+\frac{1}{2}-1}(x,\vec{k}_\perp)&=& \frac{- \sqrt{2}(k^1+ i k^2)}{1-x}\left(1+ i \frac{e_1 e_2}{8 \pi} (\vec{k}_\perp^2+B) I_4\right) \varphi, \nonumber\\
\psi^\uparrow_{-\frac{1}{2}+1}(x,\vec{k}_\perp)&=& - \sqrt{2} \Big(M- \frac{m}{x}\Big)\left(1+ i \frac{e_1 e_2}{8 \pi} (\vec{k}^2+B) I_3\right) \varphi, \nonumber\\
\label{wf_up}
\psi^\uparrow_{-\frac{1}{2}-1}(x,\vec{k}_\perp)&=& 0,
\eea
\bea
\psi^\downarrow_{+\frac{1}{2}-1}(x,\vec{k}_\perp)&=& - \sqrt{2} \Big(M- \frac{m}{x}\Big)\left(1+ i \frac{e_1 e_2}{8 \pi} (\vec{k}^2+B) I_3\right) \varphi, \nonumber\\
\psi^\downarrow_{-\frac{1}{2}+1}(x,\vec{k}_\perp)&=& \frac{- \sqrt{2}(-k^1+i k^2)}{(1-x)}\left(1+ i \frac{e_1 e_2}{8 \pi} (\vec{k}_\perp^2+B) I_4\right) \varphi, \nonumber\\
\psi^\downarrow_{-\frac{1}{2}-1}(x,\vec{k}_\perp)&=& \frac{- \sqrt{2}(k^1+i k^2)}{x(1-x)}\left(1+ i \frac{e_1 e_2}{8 \pi} (\vec{k}_\perp^2+B) I_4\right) \varphi, \nonumber\\
\psi^\downarrow_{+\frac{1}{2}+1}(x,\vec{k}_\perp)&=& 0,
\label{wf_down}
\eea
where
\be
\varphi =- e \frac{x \sqrt{1-x}}{\vec{k}_\perp^2+B},
\ee
with $B= -M^2 x(1-x) + m^2 (1-x)+\lambda^2 x$ and
\bea
I_3&=&\int_{0}^{1} \frac{1}{D'} d\alpha, \nonumber \\
I_4&=&\int_{0}^{1} \frac{\alpha}{D'} d\alpha,
\eea
with
\bea
D'=\alpha (1-\alpha)\  \vec{k}_\perp^2+\alpha \ \lambda_g^2- M^2 (1-\alpha) x (1-x)+m^2 (1-\alpha)(1-x)+ \lambda^2 (1-\alpha) x.
\eea

Using Eqs. (\ref{sivers}), (\ref{boer_mulders}), (\ref{wf_up}) and (\ref{wf_down}), we can write
\bea
f_1(x, \vec{k}_\perp)&=& \left((\frac{2 \vec{k}_\perp^2(1+x^2)}{x^2(1-x)^2}+2 \left(M-\frac{m}{x}\right)^2 \right) \varphi^2, \nonumber\\
f_{1 T}^\perp(x,\vec{k}_\perp)&=& 2 M \Big(M-\frac{m}{x}\Big) \frac{1+x}{x(1-x)} \frac{e_1 e_2}{8 \pi} (\vec{k}_\perp^2+B) \varphi^2 \frac{1}{\vec{k}_\perp^2} ln\frac{\vec{k}_\perp^2+B}{B}, \nonumber\\
h_1^\perp(x,\vec{k}_\perp)&=& M \Big(M-\frac{m}{x}\Big) \frac{1+x}{x(1-x)} \frac{e_1 e_2}{8 \pi} (\vec{k}_\perp^2+B) \varphi^2 \frac{1}{\vec{k}_\perp^2} ln\frac{\vec{k}_\perp^2+B}{B}. \label{all_distributions}
\eea

The azimuthal spin asymmetry $P_y$ can be expressed as a factorized form of the Sivers function as follows
\be
P_y= - \frac{r^1}{M}\frac{f_{1 T}^\perp}{f_1}.
\ee
In the present work, the azimuthal spin asymmetry $P_y$ and the Sivers function $f_{1 T}^\perp(x,\vec{k}_\perp)$ calculated respectively in Eqs. (\ref{final_asymmetry}) and Eq. (\ref{all_distributions}) are found to satisfy the above factorization. In the standard pQCD approach \cite{pqcd}, the azimuthal spin asymmetry factorizes as the product of Sivers function, quark distribution function, hadronic cross-section and quark fragmentation function. The cross-section in this case factorizes as the product of quark distribution function and quark fragmentation function for SIDIS at large $Q^2$.

\begin{figure}
\minipage{0.42\textwidth}
    \includegraphics[width=7cm ,angle=360]{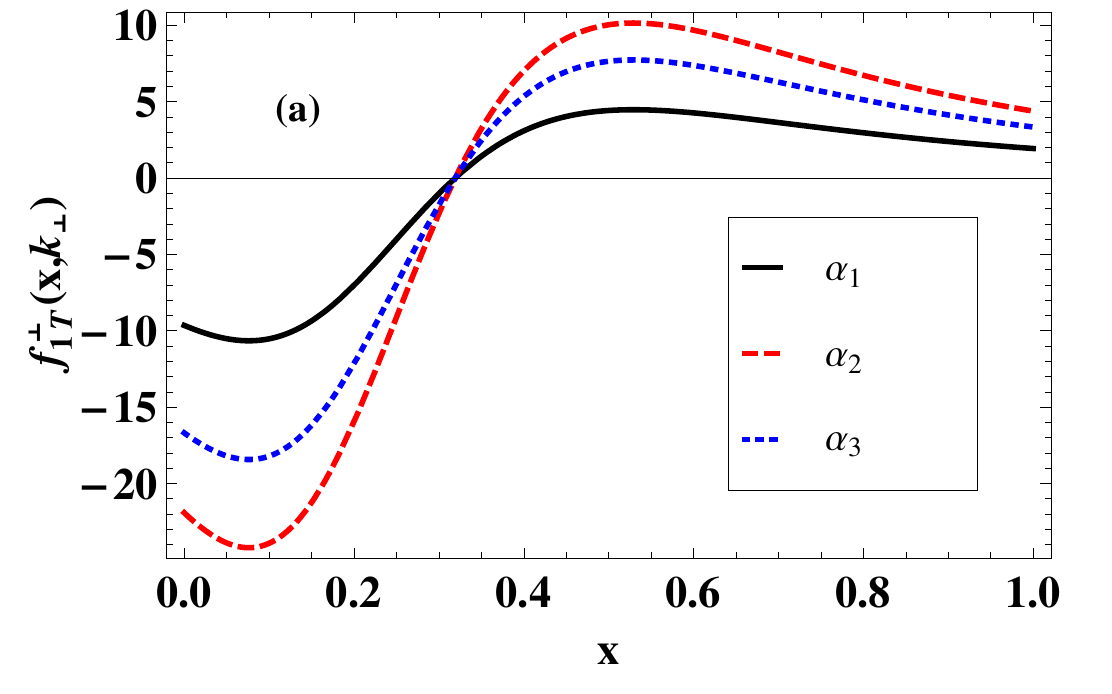}
  \endminipage\hfill
  \minipage{0.42\textwidth}
  \includegraphics[width=7cm ,angle=360]{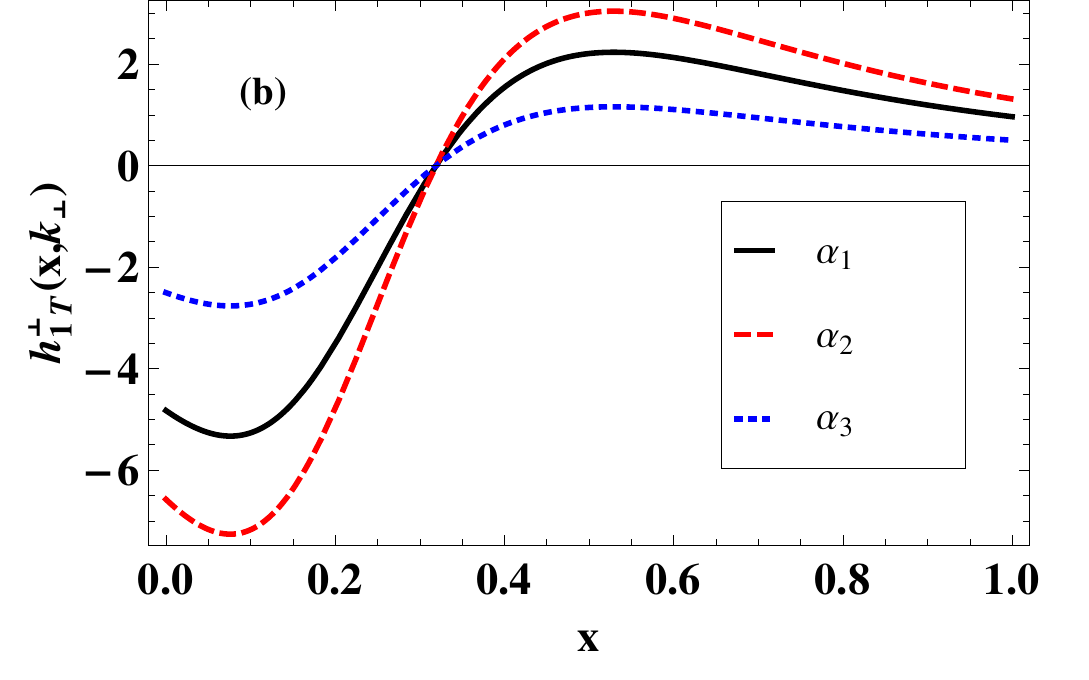}
\endminipage\hfill
\caption{Results for Sivers and Boer-Mulders function for spin-1 diquark model.}
\label{qed}
\end{figure}
In Fig. \ref{qed} we  have presented the results for Sivers and Boer-Mulders distribution functions as a function of $x$ for the fixed and running coupling constants $\alpha_1$, $\alpha_2$ and $\alpha_3$. The value of $\vec{k}_\perp$ is taken to be 0.1. We find that both Sivers and Boer-Mulders distribution functions are overall negative for $x \lesssim 0.3$ but change signs for $x \gtrsim 0.3$.  Similar behaviour has already observed in the relation for asymmetry in Eq. (\ref{final_asymmetry}). An overall negative sign for the Sivers and Boer-Mulders distribution functions has already been obtained \cite{sivers_hwang,brodsky,boer}. This occurrence of node in the present model is because of the term $\Big(M-\frac{m}{x}\Big)$ which clearly changes sign around $x \sim 0.3$ at the given value of masses in Section IV. The possibility of a node in the $x$-dependence of the Sivers and Qiu-Sterman (QS) function $T(x,S_\perp)$, related to each other by direct proportion $f_{1T}^{\perp (1)}(x)= \frac{g}{2M} T(x,S_\perp)$ \cite{boer2,sterman1,sterman2}, has also been discussed if the light cone integral is considered over the gluonic field strength \cite{grenier}. Some other model calculations \cite{zhunlu,vento} have also shown the nodes for the up and down quark Sivers function.

Calculations have also been done in the pQCD \cite{pqcd1,pqcd2} where Sivers asymmetry is shown to be related  to the QS function by considering a point like and a dipolar form factor for the interaction between the nucleon, the quark and the spectator scalar and axial-vector diquark respectively. In Ref. \cite{bacheta1,bacheta2}, the diquark model has been considered by taking the  diquark as an axial-vector rather than a vector as considered in the present work. They have also considered the longitudinal polarization vector in addition to the transverse polarization vectors. The result obtained for Sivers and Boer-Mulders distribution functions have a sign difference between them but in our calculations no sign difference has been predicted. In addition to this, the overlap representation  of Sivers and Boer-Mulders in Ref. \cite{bacheta1,bacheta2} is an assumption as it is not known a priori if the FSI operator can be isolated and the FSI operator is same for all functions and all types of diquarks. However, in our calculations the Sivers and Boer-Mulders distribution functions have been explicitly defined from the overlap representation of wave functions. This can be substantiated by the results of several future experiments that are contemplating the possibility of performing the high precision measurements over a wide $x$ region.

\section{conclusions}
We have calculated the single spin asymmetry (SSA) in the semi-inclusive deep inelastic scattering (SIDIS) process $\gamma^* p \rightarrow q (qq)_1$ using the model of light front wave functions (LFWFs) consisting of a spin-$\frac{1}{2}$ proton of mass $M$ as a composite of a spin-$\frac{1}{2}$ fermion of mass $m$ and a spin-1 vector boson of mass $\lambda$ which lies in the framework of QED. The amplitudes for the SIDIS process are computed at both the tree and one-loop level. SSAs require a phase difference between two amplitudes coupling the target with opposite spins to the same final state. The interference between these amplitudes leads to the SSA. Final-state interaction (FSI) from the gluon exchange between the outgoing quark and the target spectator system has been used to explain the observed results for the azimuthal SSAs of the proton. In particular, we have studied the dependence of transverse azimuthal spin asymmetry  ($\mathcal{P}_y$) as well as the longitudinal asymmetry  ($K \mathcal{P}_y$)  on light cone momentum fraction $\Delta$ and momentum carried by the outgoing quark $r_\perp$. The implications of such a model have been investigated in detail by considering fixed and running coupling constants. It is found that the transverse as well as the longitudinal asymmetries change sign at $\Delta \sim 0.3$ as the value of the $\Delta$ increases which may be possibly due to the $q\bar{q}$ pairs being readily produced at this momentum fraction which give rise to sea quarks affecting the sign of asymmetry at this point. Several experiments are contemplating the possibility of performing the high precision measurements over a wide $x$ region in the near future. Further, $\mathcal{P}_y$ and $K \mathcal{P}_y$ do not change sign when we fix $\Delta$ and vary the quark transverse momentum of the outgoing quark relative to photon direction$r_\perp$ and decreases as $\alpha_1 \ \Delta \ \frac{1}{r_\perp^2} ln \ r_\perp^2$. The approach presented here can also be applicable to other hadronic inclusive reactions.

The FSIs also produce a complex phase which can be included in the LFWFs to calculate the Sivers and Boer-Mulders distribution functions of the nucleon. We have also computed the Sivers and Boer-Mulders distribution function by inducing the spin-dependent complex phases to the LFWFs. Sivers function is given by the overlap of the wave functions having opposite proton spin states but the same quark spin state whereas the Boer-Mulders function has the same proton spin state but opposite quark spin state. We find an overall negative sign for both Sivers and Boer-Mulders distribution functions when $x \lesssim 0.3$. However, the signs reverse for $x \gtrsim 0.3$. The possibility of a node is also in agreement with a few other model calculations and can be related to Qiu-Sterman function.

\section{acknowledgements}
Authors would like to thank S. J. Brodsky and O. V. Teryaev for helpful discussions. H. D. would like to thank Department of Science and Technology (Ref No. SB/S2/HEP-004/2013), Government of India, for financial support.

\end{document}